\documentclass[sigplan,screen,nonacm]{acmart}

\AtBeginDocument{%
  }

\usepackage{amsmath,amsfonts}
\usepackage{algorithmic}
\usepackage{algorithm}
\usepackage{graphicx}
\usepackage{textcomp}
\usepackage{xcolor}
\usepackage{url}
\usepackage{array}
\newcolumntype{L}[1]{>{\raggedright\arraybackslash}p{#1}}

\begin{document}

\title{AFD-Ledger: Deployment Provisioning for Attention--FFN Disaggregation}

\author{Chengyu Qiu}
\authornote{Work done during Chengyu Qiu's internship at Meituan.}
\affiliation{%
  \institution{Tsinghua University}
  \city{Beijing}
  \country{China}}
\author{Xiao Fu}
\affiliation{%
  \institution{Meituan}
  \city{Beijing}
  \country{China}}
\author{Fengcun Li}
\affiliation{%
  \institution{Meituan}
  \city{Beijing}
  \country{China}}
\author{Yulei Qian}
\affiliation{%
  \institution{Meituan}
  \city{Beijing}
  \country{China}}
\author{Yuchen Xie}
\affiliation{%
  \institution{Meituan}
  \city{Beijing}
  \country{China}}
\author{Xunliang Cai}
\affiliation{%
  \institution{Meituan}
  \city{Beijing}
  \country{China}}
\author{Yingdi Shan}
\affiliation{%
  \institution{Tsinghua University}
  \city{Beijing}
  \country{China}}
\author{Yongwei Wu}
\affiliation{%
  \institution{Tsinghua University}
  \city{Beijing}
  \country{China}}
\author{Mingxing Zhang}
\correspondingauthor
\affiliation{%
  \institution{Tsinghua University}
  \city{Beijing}
  \country{China}}

\begin{abstract}

Attention--Feed-Forward Network (FFN) Disaggregation (AFD) is emerging as a
promising architecture for serving Mixture-of-Experts (MoE) language models.
While existing AFD systems improve the efficiency of disaggregated execution,
they leave a deployment question unanswered: under the same model, workload,
time-per-output-token (TPOT) service-level objective (SLO), hardware budget,
hardware catalog, and runtime capabilities, does AFD provide higher throughput
than the best collocated deployment? Answering this question requires jointly
optimizing hardware assignment and deployment organization for both
architectures, making exhaustive provisioning prohibitively expensive. We
present \emph{AFD-Ledger}, an offline analytical provisioning system that
independently provisions AFD and collocated deployments using an analytical
execution model and an evaluation-bounded hardware search. Across deployment
spaces where exhaustive provisioning is feasible, AFD-Ledger reduces complete
deployment evaluations by 68.8\%--83.5\% while still recovering the globally
optimal deployment. On three physical LongCat~2.0 deployments, it preserves
the correct architecture decision while predicting AFD-to-collocated
throughput within 6.6\%--9.6\% of measurement. Using this validated framework,
we show that homogeneous AFD improves fixed-budget throughput in only a
minority of the studied settings, heterogeneous AFD requires deployment-level
hardware complementarity rather than heuristic device selection, and
role-specific hardware improvements matter primarily when they enable better
deployment organizations by crossing deployment capability--price boundaries.

\end{abstract}

\maketitle

\section{Introduction}
\label{sec:intro}

Large language models (LLMs) are increasingly built as
Mixture-of-Experts (MoE) models. Recent frontier and production models,
including the LongCat, DeepSeek, and Qwen families, increase model capacity by
routing each token to only a small subset of feed-forward networks (FFNs),
called experts~\cite{LongCatFlash2025,DeepSeekV32,Qwen3Technical}. Because only the selected experts execute for each
token, MoE reduces the computation activated during inference and makes sparse
expert execution an important target for high-throughput serving
systems~\cite{GShard2020,SwitchTransformer2021,DeepSpeedMoE2022}.

Serving an MoE model, however, involves more than running sparse FFNs quickly.
Each decode step follows two paths with different bottlenecks. The
attention/key--value (KV) path carries request state: every active request
occupies KV cache memory, and every generated token reads that state. It is
therefore limited by memory capacity, memory bandwidth, and the number of
requests that can stay active. The routed-FFN path processes the experts
selected by each token. It benefits from larger expert batches and efficient
computation and communication. A deployment must support both paths while
meeting a time-per-output-token (TPOT) service-level objective (SLO). We call
the maximum sustained output-token rate that satisfies this target
\emph{SLO-feasible decode throughput}.

Attention--FFN Disaggregation (AFD) is a natural response to this asymmetry.
A collocated deployment runs attention and routed FFNs on the same worker pool.
AFD instead assigns request state and attention execution to
attention workers, and delegates routed expert computation to a separate
pool of FFN workers. At each MoE layer, an attention worker sends token
activations to the FFN pool and receives the routed expert outputs before
decoding continues.

The separation allows each path to be provisioned around its own bottleneck. Many attention workers can feed a smaller FFN pool, giving
each expert a larger batch and potentially higher Model FLOPs Utilization
(MFU). Removing expert weights from attention workers also leaves more memory
for KV cache. The two pools may further use different parallelism, hardware,
scheduling, and failure-recovery policies.

\begin{figure*}[t]
\centering
\includegraphics[width=0.92\textwidth]{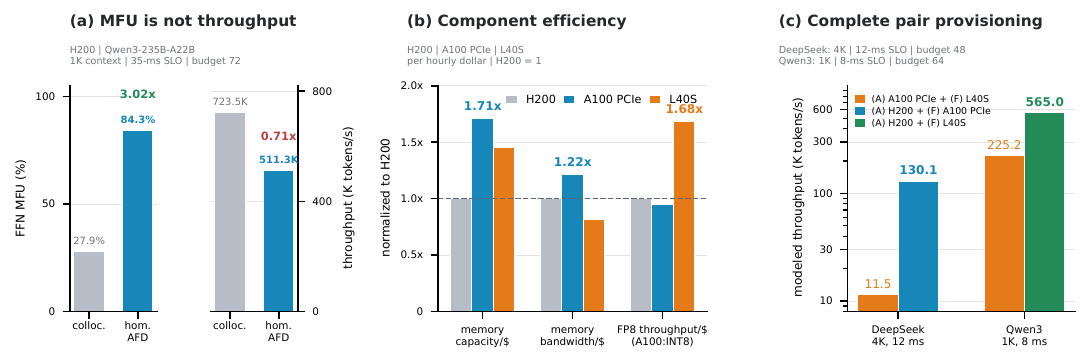}
\caption{Component-level signals and complete AFD provisioning. Prices are hourly costs~\cite{RunPodPricing2026}. Ordered pair labels name the
attention device with (A) and the FFN device with (F).}
\label{fig:teaser}
\end{figure*}

These opportunities have motivated a rapidly growing line of AFD systems that improve FFN execution, communication, scaling, dynamic scheduling, elasticity, failure recovery, or heterogeneous devices~\cite{MegaScaleInfer2025,Step32025,xiao2025xdeepserve,zhang2025janus,pan2025efficient,EaaS2025,fastafd2026}. Taken together, these systems establish that AFD can improve several important
parts of MoE serving. Their measurements, however, answer different questions.
FFN MFU describes how efficiently the expert workers execute; KV capacity
describes how many requests fit on the attention workers; heterogeneous cost
efficiency depends on the complete device mix; and elasticity and failure
isolation describe behavior over time. A gain in any one metric does not, by
itself, determine how many tokens the complete deployment serves under a fixed
budget and TPOT SLO.

This distinction motivates our paper. We do not revisit whether the individual
AFD mechanisms are useful. Instead, we ask when their benefits add up to higher
complete-deployment throughput. Our central thesis is that a local AFD gain
becomes a deployment-throughput gain only when it survives full-budget
accounting.

To see why the accounting can change the answer, consider an operator with a
fixed GPU budget. In a collocated deployment, every GPU holds request state and
executes both attention and routed FFNs. In AFD, some of those GPUs become
FFN-only workers. They may execute routed experts much more efficiently, and
removing expert weights may let each remaining attention worker hold more
requests. But the FFN-only GPUs no longer contribute KV-cache capacity. We
call the request capacity given up to create this pool the
\emph{request-bearing-capacity tax}. Homogeneous AFD improves total throughput
only when its memory and execution gains repay this tax. Heterogeneous AFD has an additional opportunity because the two
roles can use devices with different capacity, performance, and price.

We make this tradeoff quantitative for an operator provisioning a new,
saturated decode service. Given a fixed model, workload, TPOT SLO, total
budget, hardware catalog, and available runtime mechanisms, we compare the
best AFD deployment with the best collocated deployment. Each side may
optimize its own hardware assignment, parallelism, batch size, replication,
and worker organization. This comparison complements mechanism-oriented
evaluations, which appropriately hold many of these choices fixed to isolate a
new kernel, communication scheme, or scheduler. Reuse of installed hardware,
elasticity, and failure isolation remain valuable objectives, but are outside
our steady-state throughput comparison.

Figure~\ref{fig:teaser} illustrates two representative examples. The homogeneous case shows why a local FFN gain may not repay this
request-serving-capacity cost. Figure~\ref{fig:teaser}(a) compares two deployments that both use 72 H200 GPUs. Although AFD raises FFN MFU from 27.9\% to 84.3\%, the fully provisioned deployment reaches only \(0.71\times\) the throughput of the best collocated deployment. The expert pool is
far more efficient, but that local improvement does not offset the capacity
and execution costs elsewhere in the deployment.

A heterogeneous example shows the same issue from a different angle: the two
device types must work well as a pair. A natural heuristic is to use a
memory-efficient device for attention/KV workers and a compute-efficient
device for FFN workers. In the catalog in Figure~\ref{fig:teaser}(b), this heuristic selects
A100 PCIe for attention and L40S for FFNs. Yet under the corresponding fixed
budget and TPOT SLO, this AFD pair reaches only \(0.09\times\) and \(0.40\times\) the modeled
throughput of another less obvious pair, as shown in Figure~\ref{fig:teaser}(c). This result shows that the
best-looking device for each role in isolation need not produce the best
complete deployment.

The two examples turn the deployment-performance question into a systems
search problem over hardware assignment and deployment organization. Existing analytical provisioning systems demonstrate that deployment performance can be evaluated without exhaustively implementing every candidate\cite{AnalyticalAFD2026,Frontier2026,HowFarDisaggregation2026}. However, they typically assume that the hardware pool has already been determined before provisioning begins. As a result, their provisioning decisions and
deployment insights are specific to the chosen hardware rather than the
hardware catalog available to an operator. Once hardware selection also becomes
part of provisioning, every hardware assignment induces a complete deployment
search, making exhaustive exploration expensive.

We present \emph{AFD-Ledger}, an offline analysis and provisioning system for
this search. Given the deployment conditions above, AFD-Ledger explores the
AFD and collocated spaces independently. It first uses an analytical model to
estimate complete plans across hardware assignments, then spends detailed
planning effort on only a small set of promising assignments. Its output is
not a general verdict on AFD, nor another evaluation under a
fixed hardware pool: it identifies whether AFD improves throughput in
the given setting and which workload, SLO, budget, and hardware conditions enable that
improvement.

We evaluate AFD-Ledger from two perspectives. Against exhaustive analytical provisioning, it reduces complete deployment evaluations by
68.8\%--83.5\% while still recovering the global optimum. Against three
physical LongCat deployments, it predicts the correct architecture decision with throughput errors of only 6.6\%--9.6\%. This observed error is not a universal confidence bound; we
therefore report smaller retrospective gaps as near ties or sensitivity
results.

Using this validated framework, we obtain three main findings.

\emph{First, homogeneous AFD provides higher fixed-budget throughput in only a
minority of the studied settings.} Across 36 homogeneous deployment settings over our studied hardware
catalog, AFD-Ledger selects AFD in 7 cases. It wins only when
expert-memory removal, larger batches, and lower TPOT are sufficient to repay
the request-bearing-capacity tax.

\emph{Second, heterogeneous AFD can provide larger gains, but only with the
right hardware complementarity.} The opportunity does not follow simply from
assigning a memory-efficient device to attention and a compute-efficient
device to FFNs. Across the studied conditions, the throughput ratio of the
same AFD hardware pair relative to collocation ranges from \(0.483\times\) to
\(1.815\times\) as the TPOT target and budget change. A useful pair must
jointly provide the right memory capacity, bandwidth, compute, communication,
price, and worker ratio after the complete deployment is provisioned.

\emph{Third, role-specific hardware improvements matter most when they unlock
a better deployment organization.} Deployment choices are discrete: a device
improvement may have little effect until it makes a new replica layout or
attention--FFN worker ratio feasible. In a representative DeepSeek deployment,
crossing such a capability--price boundary increases throughput by \(19.59\times\).

In summary, this paper makes three contributions:

\begin{itemize}
    \item We separate the major benefits commonly grouped under AFD and define
    a full-deployment accounting objective for its performance claim: the best
    SLO-feasible decode throughput under a common deployment specification.

    \item We present AFD-Ledger, an analytical provisioning system that
    extends existing model-based provisioning from fixed hardware configurations to complete deployment provisioning over hardware catalogs, while
    fully evaluating only a small set of promising hardware assignments.

    \item We quantify when AFD's local and role-specific benefits translate into
    deployment-level throughput. In our studied space, homogeneous AFD is
    selected in only 7 of 36 settings, whereas heterogeneous AFD can provide larger
    gains but requires workload-, SLO-, and budget-specific hardware
    complementarity. We further derive practical principles for AFD hardware
    selection and design.
\end{itemize}

\section{Accounting for AFD Benefits at Deployment Scale}
\label{sec:background}

Recall that AFD replaces one collocated worker pool with an attention/KV pool
and an FFN pool; a complete deployment must provision and connect both. We now
make the resulting accounting concrete. We first walk through one homogeneous
and one heterogeneous example, reproducing the reported AFD gains analytically and separating that evidence from our full reprovisioning. We then use those examples to derive the fixed-budget ledger
and formal comparison used in the rest of the paper.

\subsection{A Homogeneous Example}

FastAFD evaluates Qwen3-235B serving at an 8K context on a GB200 NVL72 cluster~\cite{fastafd2026}. It reports two deployments: an expert parallelism (EP) width of four for the collocated deployment and an AFD deployment that uses seven four-GPU attention nodes and one four-GPU FFN node (28 attention and 4 FFN GPUs). The reported throughput per GPU increases from 1,781 to 2,518 output tokens/s, corresponding to a measured \(2{,}518/1{,}781=1.414\times\) gain. We first reconstruct this reported deployment analytically to understand where the improvement comes from before asking whether the deployment itself is optimal.

\begin{figure}[t]
\centering
\includegraphics[width=\linewidth]{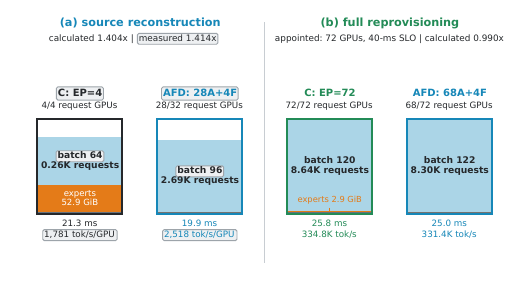}
\caption{FastAFD Qwen3-235B 8K accounting. Gray, orange, and
blue denote attention weights, expert weights, and occupied KV cache. Gray-backed labels identify source-reported data/results.}
\label{fig:fastafd_mechanism}
\end{figure}

Throughput during steady-state decode depends on only two quantities: how many requests can decode concurrently and how long one decoding step (TPOT) takes. We therefore reconstruct these two quantities separately.

We begin with concurrency. Each GB200 provides 185.0~GiB of memory available to model state. For Qwen3-235B, attention and routing weights occupy 3.4~GiB per GPU, while the model contains 211.5~GiB of expert weights. With EP=4, the experts are divided across four GPUs, so every request-serving GPU stores \( 3.4 + \frac{211.5}{4}=56.3\ \text{GiB} \) of model weights, leaving \( 185.0-56.3=128.7\ \text{GiB} \) for KV cache. The published model configuration reports a bfloat16 (BF16) KV footprint of 188~KiB per token. At an 8K context, each resident request therefore occupies \( 188\ \text{KiB}\times8192 =1.469\ \text{GiB}. \) The remaining KV memory can therefore hold \( 128.7/1.469\approx87 \) resident requests per GPU. The reported AFD deployment moves all expert weights onto four dedicated FFN workers. Each attention worker therefore stores only the 3.4~GiB attention and routing weights, leaving \(185.0-3.4=181.6\ \text{GiB}\) for KV cache. This corresponds to \( 181.6/1.469\approx123 \) resident requests. The reported batch size of 96 therefore fits comfortably within the available memory, increasing the feasible resident requests per attention GPU from about 87 to 123 (Figure~\ref{fig:fastafd_mechanism}(a)).

We next reconstruct TPOT. For every decoded token, the published operator statistics specify both the required floating-point operations and high-bandwidth memory (HBM) traffic. Stage latency is estimated as the slower of computation and memory transfer. For the reported EP=4 deployment at batch 64, attention requires 0.507~ms of compute but 13.111~ms of HBM transfer based on peak compute and HBM bandwidth, making it memory-bound. The reconstructed FFN stage requires 7.300~ms, while dispatch and combine each add 0.438~ms over the peak 900~GB/s interconnect. Executing these stages sequentially gives a TPOT of \(13.111+0.438+7.300+0.438=21.287\)~ms.

For AFD, the same stage-level model additionally accounts for pipelined
execution between attention and FFN workers. The resident batch is divided into microbatches, whose number equals the pipeline depth\cite{MegaScaleInfer2025}. The reported deployment uses pipeline depth two, executing the 96-request batch as two 48-request
microbatches. Each microbatch requires 9.947~ms for attention, 8.164~ms for FFNs, and 0.329~ms for each communication direction. The resulting pipeline is attention-bound, giving a TPOT of \(9.947\times2=19.894\)~ms.

The reconstructed throughput now follows directly. Steady-state throughput is proportional to the number of request-serving GPUs, the resident batch size, and inversely proportional to TPOT. The reconstructed deployment
therefore achieves \( \frac{28}{32} \times \frac{96}{64} \times \frac{21.287}{19.894} = 1.404, \) within 0.7\% of the reported \(2518/1781=1.414\times\) improvement, confirming that the
reported gain follows directly from larger resident batches and lower TPOT.

Having reproduced the reported result, we now ask a different question. The published comparison fixes the collocated deployment to EP=4. Provisioning instead asks whether AFD remains the better deployment when both AFD and collocated deployment are allowed to choose their best organization.

Figure~\ref{fig:fastafd_mechanism}(b) therefore reprovisions both architecture families under the same deployment specification: 72 GB200 GPUs and a 40-ms TPOT SLO. Using exactly the reconstruction procedure above, we exhaustively enumerate all legal collocated EP configurations together with all feasible AFD attention/FFN splits, batching strategies, and pipeline depths.

The best AFD deployment changes from the reported 28+4 organization to 68+4. The larger attention pool supports batch 122 with 8,296 concurrent requests. Its reconstructed TPOT is 25.036~ms, yielding 331.4K output tokens/s. The best collocated deployment instead widens expert parallelism from EP=4 to EP=72, reducing expert memory sufficiently to support batch 120 across all 72 GPUs. The resulting deployment reaches 8,640 concurrent requests with a 25.810-ms TPOT, corresponding to 334.8K output tokens/s.

The two deployments therefore differ by only \( 331.4/334.8=0.990\) , which falls within the near-tie range established by
our physical validation. This does not contradict FastAFD's conclusion: the
reported comparison shows that its mechanisms improve the selected EP=4
deployment, whereas reprovisioning asks whether they remain sufficient once
both deployment families are independently optimized.

\subsection{A Heterogeneous Example}

The homogeneous example considered only one GPU type. Heterogeneous AFD introduces an additional degree of freedom by allowing attention and FFN workers to use different devices, making hardware assignment part of the provisioning problem.

MegaScale-Infer studies this opportunity using DBRX. Following the natural hardware-selection heuristic illustrated in Section~\ref{sec:intro}, their work selects H20 GPUs for attention and L40S GPUs for FFNs among its hardware catalog (L20, H800, A800, H20, L40S)~\cite{MegaScaleInfer2025}. Under the reported workload, the paper measures a \(1.574\times\) throughput improvement over an H20 collocated baseline. Because the deployment organization is not reported, we reproduce the comparison using a source-aligned proxy with the reported model, workload, and hardware pair, appointing only the missing deployment parameters required for analytical evaluation.

\begin{figure}[t]
\centering
\includegraphics[width=\linewidth]{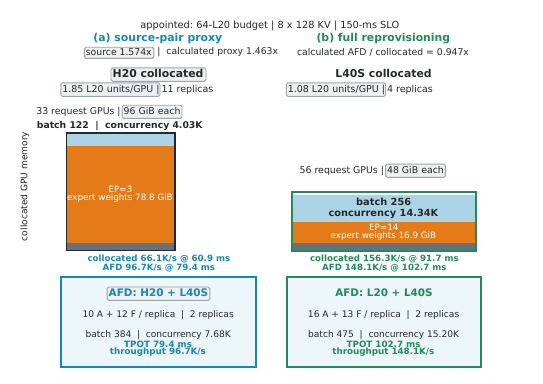}
\caption{MegaScale-Infer DBRX proxy analysis at the reported 571-token context
and 150-ms TPOT SLO, with an appointed 64-L20-equivalent budget. Gray-backed labels identify source-reported data/results.}
\label{fig:megascale_mechanism}
\end{figure}

Applying the analytical execution model developed in the previous example, the appointed H20 collocated deployment provisions 33 request-serving GPUs. The heterogeneous organization instead provisions 20 H20 attention GPUs together with 24 L40S FFN GPUs. Although fewer GPUs retain requests, replacing H20 with L40S increases aggregate FFN compute from 4.88 to 8.69~PFLOP/s, a \(1.78\times\) improvement under essentially the same budget.

Using the same throughput accounting, the collocated proxy reaches 66.1K output tokens/s, while the heterogeneous deployment reaches 96.7K output tokens/s. The improvement comes from two complementary effects: removing expert weights increases KV capacity on the attention workers, while the stronger L40S expert pool absorbs the larger routed workload without proportionally increasing TPOT. The reconstructed \(1.463\times\) throughput gain remains within 7.1\% of the paper's reported \(1.574\times\) result.

We next enlarge the provisioning problem by allowing both architecture families to choose freely from the complete hardware catalog. For every hardware assignment, we optimize deployment organization, batching, parallelism, replication, and worker allocation.

The best collocated deployment changes to L40S because its lower price permits substantially more request-serving GPUs under the same budget, reaching 156.3K output tokens/s. The best heterogeneous deployment instead selects L20 for attention and L40S for FFNs, balancing KV capacity with expert throughput, and reaches 148.1K output tokens/s.

The reported comparison therefore answers a different question from deployment provisioning. It demonstrates that one appointed H20+L40S deployment outperforms one appointed H20 baseline. Once hardware selection itself becomes part of provisioning, the throughput ratio changes to
\(
148.1/156.3=0.947,
\)
which falls within the near-tie range established by our physical validation.

\subsection{A Fixed-Budget Performance Ledger}

The two examples above follow the same accounting principle: under a fixed deployment budget, every benefit created by AFD must ultimately compensate for the resources consumed to create it. For homogeneous AFD, this cost is the request-bearing-capacity tax; for heterogeneous AFD, hardware specialization also changes how many request-serving workers the budget can afford.

In steady-state decode, if a deployment has \(N_r\) request-serving GPUs, each holding \(b\) resident requests, and achieves a TPOT of \(T\), then its output-token throughput is
\begin{equation}
    \Theta
    =
    \frac{N_r b}{T}.
\end{equation}

Comparing an AFD deployment with a collocated deployment therefore gives
\begin{equation}
    \frac{\Theta_A}{\Theta_C}
    =
    \underbrace{\frac{N_{r,A}}{N_{r,C}}}_{\text{request-serving GPUs}}
    \cdot
    \underbrace{\frac{b_A}{b_C}}_{\text{requests per GPU}}
    \cdot
    \underbrace{\frac{T_C}{T_A}}_{\text{TPOT}}.
    \label{eq:throughput-ratio}
\end{equation}

Applying Eq.~\ref{eq:throughput-ratio} to the FastAFD example reproduces both comparisons. The reported deployment gives
\(
\frac{28}{32}\cdot\frac{96}{64}\cdot\frac{21.287}{19.894}
\approx1.404,
\)
matching the reported \(1.414\times\) gain. After independently reprovisioning both deployment families under the same 72-GPU budget,
\(
\frac{68}{72}\cdot\frac{122}{120}\cdot\frac{25.810}{25.036}
\approx0.990.
\)
The accounting is unchanged; only the deployment organization differs.

\subsection{Comparing the Best Plans Under the Same Conditions}
\label{sec:comparison-contract}

The previous sections establish our analytical execution model, which estimates the SLO-feasible throughput of any deployment analytically. Provisioning then asks a second question: which deployment should be compared?

We formulate this comparison by fixing the deployment specification and allowing each architecture to search for its best valid configuration.

We write the conditions fixed before this search as

\begin{equation}
    \Omega = (M, W, \tau, B, \mathcal{H}, \Gamma),
    \label{eq:deployment-specification}
\end{equation}
where \(M\), \(W\), \(\tau\), and \(B\) denote the model, workload, TPOT SLO, and budget; \(\mathcal H\) is the hardware catalog; and \(\Gamma\) captures shared deployment assumptions (e.g., network topology, supported kernels, parallelism, and memory reservations).

For one deployment \(d\), let \(\mathcal F_d\) denote the feasible resident concurrencies and \(T_d(C)\) the corresponding TPOT. Its score is the SLO-feasible decode throughput:

\begin{equation}
    \Theta(d;\tau)
    =
    \max_{C\in\mathcal{F}_d}
    \left\{
        \frac{C}{T_d(C)}
        \ \middle|\
        T_d(C)\leq\tau
    \right\}.
    \label{eq:slo-feasible-throughput}
\end{equation}

Provisioning compares the best collocated and AFD deployments:

\begin{equation}
    \begin{aligned}
        \Theta_C^\star(\Omega)
        &=
        \max_{d\in\mathcal{D}_{C}(\Omega)}\Theta(d;\tau),\\
        \Theta_A^\star(\Omega)
        &=
        \max_{d\in\mathcal{D}_{A}(\Omega)}\Theta(d;\tau).
    \end{aligned}
    \label{eq:family-optima}
\end{equation}
where \(\mathcal{D}_{C}(\Omega)\) and \(\mathcal{D}_{A}(\Omega)\) are the legal collocated and AFD plans

This objective rewards both higher concurrency and lower TPOT while enforcing the TPOT SLO. As before, analytical differences smaller than the error observed in physical validation are reported as near ties rather than conclusive wins.

\subsection{Relating the Accounting to Earlier AFD Results}

Earlier AFD papers typically evaluate whether a proposed mechanism improves a selected deployment, whereas our objective asks whether it changes the best deployment under a common deployment specification. We therefore retain each paper's reported comparison before introducing any alternative deployment. When the deployment organization is reported (e.g., FastAFD), we reconstruct that exact deployment and then independently reprovision both architecture families. When it is omitted (e.g., MegaScale-Infer), we perform a source-aligned proxy analysis using explicitly stated assumptions before reprovisioning.

Our retrospective analysis targets papers whose primary claim is deployment-level throughput improvement over collocated serving. Among existing AFD systems, only FastAFD and MegaScale-Infer provide the direct deployment-level comparisons required for reprovisioning; other systems pursue different objectives or do not report comparable deployment results~\cite{MegaScaleInfer2025,fastafd2026,Step32025,xiao2025xdeepserve,pan2025efficient,RevealingAFD2026,zhang2025janus,EaaS2025}.

\begin{figure}[t]
\centering
\includegraphics[width=\linewidth]{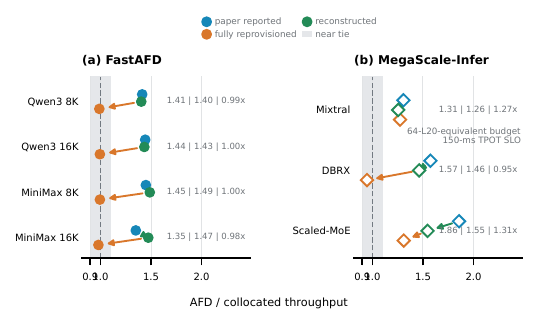}
\caption{Reported, reconstructed and fully reprovisioned AFD-to-collocated throughput ratios. Filled markers denote reconstruction and 
open markers denote proxy. The gray band marks the near-tie range set observed in physical validation.}
\label{fig:comparison_protocol_effect}
\end{figure}

Figure~\ref{fig:comparison_protocol_effect} summarizes the protocol. Reconstructing the reported deployments preserves their original conclusions, confirming that our analytical model reproduces the published comparisons. Reprovisioning instead enlarges the deployment search space. Several reported advantages shrink to near ties or become sensitive to additional deployment choices, showing that deployment-level conclusions depend on the configurations included in the comparison rather than only on the proposed AFD mechanism.

\subsection{The Remaining Provisioning Challenges}
\label{sec:challenges}

The retrospective examples above involve only a few candidate deployments and
can therefore be analyzed manually. Practical provisioning, however, must
consider an entire hardware catalog rather than a handful of appointed
configurations.

Existing analytical provisioning systems already avoid implementing every
candidate deployment, but they assume that the hardware pool is fixed before
provisioning begins~\cite{Frontier2026,
HowFarDisaggregation2026,AnalyticalAFD2026}. Once hardware selection also
becomes part of provisioning, every hardware assignment induces another search
over worker allocation, batching, parallelism, replication, pipelines, and
communication under the same deployment objective, greatly enlarging the search
space.

At the same time, hardware-aware provisioning is no longer only an optimization
problem. Comparing complete AFD and collocated deployments across hardware
catalogs also reveals when AFD should be adopted and what hardware properties
make it beneficial. Solving this expanded provisioning problem efficiently is
the goal of AFD-Ledger, presented next.

\section{Design}
\label{sec:design}

We present AFD-Ledger, a provisioning system that addresses the problem introduced in the last section through a two-level provisioning strategy. it first uses inexpensive role-specific estimates to prioritize promising hardware assignments, then performs complete deployment provisioning only for the selected candidates.

\begin{figure}[t]
\centering
\includegraphics[width=\linewidth]{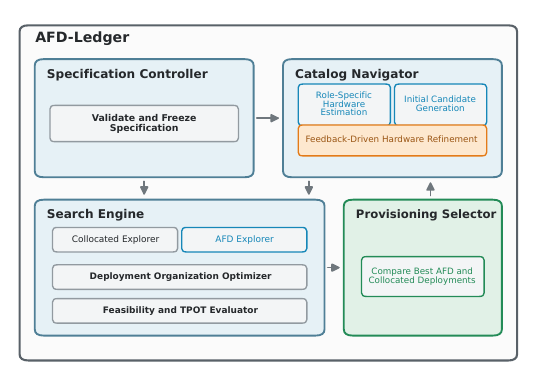}
\caption{AFD-Ledger system architecture}
\label{fig:framework}
\end{figure}

Figure~\ref{fig:framework} summarizes the workflow. The deployment specification is first fixed once. The collocated optimum is provisioned once because it depends only on the hardware catalog. For AFD, the \emph{Catalog Navigator} proposes hardware assignments, the \emph{Search Engine} fully provisions each selected assignment, and the resulting throughput is fed back to guide later exploration. Consequently, exhaustive deployment provisioning is performed only for a small fraction of hardware assignments.

\begin{algorithm}[t]
\caption{Fixed-catalog provisioning loop}
\label{alg:fixed_catalog}
\footnotesize
\begin{algorithmic}[1]
\REQUIRE Fixed deployment specification $\Omega$; assignment budget $Q$
\REQUIRE Retained width $w$; expansion width $x$; refinement-round limit $K$
\ENSURE Highest-throughput deployment found under $\Omega$
\STATE $\textsc{Validate and Freeze}(\Omega)$
\STATE $c^*\leftarrow\textsc{Provision Best Collocated}(\Omega)$
\STATE $(P_A,P_E)\leftarrow\textsc{Role-Specific Hardware Estimation}(\Omega)$
\STATE $R\leftarrow\textsc{Initial Candidate Generation}(P_A,P_E)$
\STATE $E\leftarrow\emptyset$; $A\leftarrow\emptyset$
\STATE $(E,A,Q)\leftarrow\textsc{Consistent Provisioning}
       (\Omega,R,E,A,Q)$
\STATE $k\leftarrow0$
\WHILE{$Q>0$ and $E\neq\emptyset$ and $k<K$}
    \STATE $B\leftarrow\textsc{Retain Role-Diverse Deployments}(E,w)$
    \STATE $R\leftarrow\textsc{Feedback-Driven Hardware Refinement}
           (B,P_A,P_E,A)$
    \STATE $R\leftarrow$ highest-ranked $\min\{x,Q\}$ assignments in $R$
    \IF{$R=\emptyset$} \STATE \textbf{break} \ENDIF
    \STATE $(E,A,Q)\leftarrow\textsc{Consistent Provisioning}
           (\Omega,R,E,A,Q)$
    \STATE $k\leftarrow k+1$
\ENDWHILE
\IF{$Q>0$}
    \STATE $R\leftarrow\textsc{Highest-Ranked Unseen Assignments}
           (P_A,P_E,A,Q)$
    \STATE $(E,A,Q)\leftarrow\textsc{Consistent Provisioning}
           (\Omega,R,E,A,Q)$
\ENDIF
\IF{$E=\emptyset$}
    \STATE \textbf{return} $c^*$
\ENDIF
\STATE $a^*\leftarrow$ highest-throughput deployment in $E$
\STATE \textbf{return} $\arg\max_{d\in\{c^*,a^*\}}\textsc{Throughput}(d)$
\end{algorithmic}
\end{algorithm}

Algorithm~\ref{alg:fixed_catalog} formalizes this process. The search is given a budget of at most \(Q\) hardware assignments. It first constructs an initial set of candidate pairs, provisions them completely, and records both feasible deployments (\(E\)) and attempted assignments (\(A\)). Each refinement round retains up to \(w\) high-quality but hardware-diverse deployments, generates neighboring assignments by changing one execution role, and provisions at most \(x\) new candidates. Once neighborhood refinement no longer makes progress, any remaining budget is spent on the highest-ranked unseen assignments from the entire catalog. The best AFD deployment is finally compared against the precomputed collocated optimum under the same deployment specification.

The remainder of this section focuses on the two key components of AFD-Ledger: efficient hardware exploration and consistent deployment provisioning.

\subsection{Efficient Hardware Exploration}
\label{sec:hardware_search}

Evaluating one hardware assignment requires complete deployment provisioning, making exhaustive search over all attention--expert hardware pairs prohibitively expensive. AFD-Ledger therefore separates hardware exploration from deployment provisioning. Lightweight hardware estimates determine which assignments are worth evaluating, while complete provisioning determines their actual throughput.

\subsubsection{Role-Specific Hardware Estimation}

During hardware exploration, complete deployment provisioning is unavailable. AFD-Ledger therefore estimates hardware quality independently for attention and expert execution.

Each compatible device is first checked against the minimum memory requirement for its role. It is then evaluated on several representative probe batches (1, 4, 16, 64) using the analytical execution model introduced in Section~\ref{sec:background}. The highest memory-feasible service rate becomes its role estimate. To account for hardware cost, both service rate and HBM capacity are further normalized by the device price.

These estimates intentionally ignore deployment variables such as worker allocation, replication, communication, and pipeline organization. They are used only to prioritize hardware assignments for complete provisioning rather than to predict deployment throughput.

\subsubsection{Initial Candidate Generation}

The role estimates are next combined to construct the initial hardware assignments. Rather than pairing every attention device with every expert device, AFD-Ledger first retains a small set of representative devices for each execution role, including the cheapest device, the highest-service device, the highest service-per-cost device, and the highest memory-per-cost device.

Candidate hardware pairs are ranked using
\[
\widehat R(h_A,h_E)
=
\frac{2}
{(R_A(h_A)/p_A)^{-1}
+(R_E(h_E)/p_E)^{-1}},
\]
where \(R_A\) and \(R_E\) denote the attention and expert service estimates, and \(p_A\) and \(p_E\) are the corresponding H200-normalized hardware prices. The harmonic mean favors balanced cost efficiency across the two execution roles.

The score serves only to prioritize complete provisioning and is never interpreted as deployment throughput.

\subsubsection{Feedback-Driven Hardware Refinement}

After the initial candidates have been fully provisioned, subsequent hardware exploration is guided by deployment throughput rather than analytical estimates.

AFD-Ledger first retains the highest-throughput feasible deployments while preferring hardware pairs that introduce previously unseen attention or expert devices. This role-diverse retained set prevents the search budget from being consumed by many neighboring deployments built around the same hardware.

Each retained pair then generates neighboring assignments by replacing either its attention device or its expert device while keeping the other execution role unchanged. These neighbors are ranked using the lightweight role estimates, and only the highest-ranked candidates are selected for the next round of complete provisioning. After a bounded number of refinement rounds, any remaining evaluation budget is finally allocated to the highest-ranked unevaluated hardware assignments across the entire catalog, allowing both execution roles to change simultaneously and preventing the search from remaining trapped around one local region of the hardware space.

\begin{figure}[t]
\centering
\includegraphics[width=\linewidth]{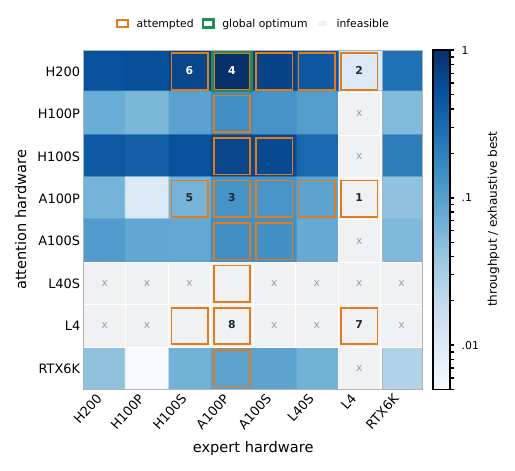}
\caption{DeepSeek-V3.2 ordered-assignment landscape for H200 SXM, H100 PCIe
and SXM, A100 PCIe and SXM, L40S, L4, and RTX PRO 6000 Server Edition at 4K
context, 12-ms TPOT, and a budget of 48 H200-equivalent units. Axis suffixes P and S denote PCIe and SXM; RTX6K denotes RTX PRO 6000 Server
Edition.}
\label{fig:pair_landscape}
\end{figure}

Figure~\ref{fig:pair_landscape} illustrates the exploration process. Although the hardware catalog contains 64 ordered assignments, AFD-Ledger provisions only 20 of them before identifying the globally optimal assignment. Early hardware estimates guide the initial exploration, while subsequent provisioning results quickly redirect the search toward better regions of the hardware space.

\subsection{Consistent Deployment Provisioning}
\label{sec:deployment_search}

The Catalog Navigator selects hardware assignments for complete evaluation. For each selected assignment, AFD-Ledger provisions the best deployment under the same deployment specification $\Omega$ and returns its SLO-feasible decode throughput. Using an identical provisioning procedure for both AFD and collocated serving ensures that any throughput difference reflects the deployment architecture rather than inconsistencies in the search process.

\subsubsection{Deployment Enumeration}

Different deployment architectures expose different search variables.
Collocated serving optimizes EP width, replication, batching, and scheduling, whereas AFD additionally optimizes the attention--expert worker allocation and pipeline depth.

For each selected hardware assignment, AFD-Ledger enumerates all feasible
deployment configurations permitted by $\Omega$, evaluates them using the
common analytical execution model, and retains only the highest-throughput
deployment. Every attempted hardware assignment consumes one evaluation
regardless of feasibility, ensuring that the exploration budget counts
complete provisioning attempts rather than successful deployments.

\subsubsection{Deployment Organization}

All deployment configurations are evaluated using the common analytical
execution model introduced in Section~\ref{sec:background}. The model performs
memory checking, TPOT estimation, schedule construction, and throughput
calculation for both collocated and AFD deployments. The only difference between the two searches is the deployment variables being enumerated.

\section{Implementation}
\label{sec:implementation}

Our implementation consists of two parts described below.

\subsection{Offline Provisioning System}

We implement AFD-Ledger as an offline Python provisioning system comprising approximately 5,700 lines of code (excluding tests). The implementation includes hardware exploration, deployment provisioning, and an exhaustive search mode used as the reference in experiments where full enumeration is tractable.

\subsection{Physical AFD Runtime}

To validate the analytical assumptions of AFD-Ledger, we separately implement a production-ready AFD runtime for LongCat~2.0~\cite{LongCat20Model2026} on top of the company's internal SGLang-FluentLLM codebase. Unlike the offline provisioning system, this runtime executes the complete AFD serving path and is used exclusively for physical validation.

Implementing the runtime requires substantially more engineering effort than the analytical provisioning system because AFD is not natively supported by the underlying serving stack. We build separate attention and expert worker pools, implement the dispatch--execution--combine execution path, develop an AFD scheduler, establish cross-pool communication, and integrate microbatch and DP-domain pipelining~\cite{xiao2025xdeepserve}. We further adapt LongCat's zero-computation expert execution to Ascend-specific operators, support Graph Engine (GE) execution, and validate full-layer numerical correctness on Ascend 910C SuperPods.

The runtime comprises more than 8,000 lines of Python and AscendC code, demonstrating that the analytical assumptions are validated against a complete end-to-end implementation rather than a simplified prototype.

\section{Evaluation}
\label{sec:evaluation}

We evaluate AFD-Ledger from three perspectives.

\textbf{Q1.} Can AFD-Ledger accurately provision complete deployments while avoiding exhaustive hardware search?

\textbf{Q2.} What deployment insights emerge from comparing fully provisioned AFD and collocated deployments?

\textbf{Q3.} Do these conclusions remain stable under future hardware and modeling assumptions?

\subsection{Experimental Setup}
\label{sec:exp_setup}

We evaluate Qwen3-235B-A22B and DeepSeek-V3.2, two representative frontier MoE models.

\begin{table}[t]
\centering
\caption{Configured commercial hardware and the two selected simulated target points interpreted in the future-hardware study. Prices
are hourly costs normalized to H200~\cite{RunPodPricing2026}. NVIDIA specifications
follow vendor documentation~\cite{NVIDIADataCenterGPUs2026}.}
\label{tab:hardware_catalog}
\footnotesize
\setlength{\tabcolsep}{1.5pt}
\begin{tabular}{L{0.28\linewidth}ccccc}
\hline
Device & Price & HBM (GiB) & BW (TB/s) & 8-bit (TFLOP/s) & Role \\
\hline
\multicolumn{6}{l}{\textit{E Catalog}} \\
H200 SXM & 1.000 & 141 & 4.80 & 1978 & General \\
H100 PCIe & 0.554 & 80 & 2.00 & 1513 & General \\
H100 SXM & 0.749 & 80 & 3.35 & 1978 & General \\
A100 PCIe & 0.331 & 80 & 1.94 & 624 (INT8) & General \\
A100 SXM & 0.387 & 80 & 2.04 & 624 (INT8) & General \\
L40S & 0.220 & 45 & 0.86 & 733 & General \\
L4 & 0.123 & 24 & 0.30 & 242 & General \\
RTX PRO 6000 Server & 0.471 & 96 & 1.79 & 936 & General \\
\multicolumn{6}{l}{\textit{S Catalog}} \\
B200 SXM & 1.641 & 180 & 8.00 & 4500 & General \\
B300 SXM & 1.933 & 269 & 7.75 & 4500 & General \\
Sim. attention target & 0.100 & 30 & 8.00 & 4500 & Attention \\
Sim. expert target & 0.020 & 7.5 & 1.50 & 200 & Expert \\
\hline
\end{tabular}
\end{table}

Table~\ref{tab:hardware_catalog} summarizes the evaluated hardware catalogs. Catalog E contains most commercial accelerators and is used throughout the main evaluation, while Catalog S extends it with Blackwell GPUs and simulated role-specialized devices for the hardware co-design study.

\begin{table*}[t]
\centering
\caption{Evaluation specifications. Budgets are hourly H200-equivalent units. Specifications
counts the deployment settings represented by a row; Pair entries give bounded
attempts/exhaustive role-compatible ordered assignments. All analytical rows use
\(Q=20\), retained width \(w=2\), expansion width \(x=6\), and at most
\(K=3\) refinement rounds. Braces denote swept values.}
\label{tab:reference_cases}
\footnotesize
\setlength{\tabcolsep}{3.4pt}
\resizebox{0.98\textwidth}{!}{
\begin{tabular}{lllcccccl}
\hline
Case or suite & Model & Catalog & Input & TPOT SLO (ms) & Budget & Specifications &
Pairs & Evaluation use \\
\hline
\multicolumn{9}{l}{\textit{Five reference cases}} \\
D/E & DeepSeek & E & 4K & 12 & 48 & 1 & 20/64 &
search, existing result \\
Q/E & Qwen & E & 1K & 8 & 64 & 1 & 20/64 &
search, existing result \\
D/S8 & DeepSeek & S & 4K & 8 & 48 & 1 & 20/121 &
search, future hardware \\
D/S12 & DeepSeek & S & 4K & 12 & 48 & 1 & 20/121 &
search, future hardware, sensitivity \\
Q/S & Qwen & S & 1K & 10 & 64 & 1 & 20/121 &
search, future hardware \\
\multicolumn{9}{l}{\textit{Deployment-behavior suites}} \\
E-grid & DeepSeek & E & 4K & \(\{8,10,12,14,16,20\}\) &
\(\{32,48,64\}\) & 18 & 20/64 & homogeneous, decisions, ablation \\
E-grid & Qwen & E & 1K & \(\{6,8,10,12,15,20\}\) &
\(\{32,48,64\}\) & 18 & 20/64 & homogeneous, decisions, ablation \\
S-sweep & DeepSeek & S & 4K & \(\{8,10,12,14,16,20\}\) &
48 & 6 & 20/121 & future-hardware SLO sweep \\
S-sweep & Qwen & S & 1K & \(\{6,8,10,12,15,20\}\) &
64 & 6 & 20/121 & future-hardware SLO sweep \\
\multicolumn{9}{l}{\textit{Physical validation}} \\
LongCat & LongCat~2.0 & Ascend 910C & \(\{32\text{K},64\text{K},120\text{K}\}\) &
-- & 160 cards & 3 & -- & analytical/physical comparison \\
\hline
\end{tabular}}
\end{table*}

Table~\ref{tab:reference_cases} summarizes the evaluation suites, including five reference cases used throughout the paper, parameter sweeps over TPOT and hardware budgets, and physical validation on LongCat~2.0.

\subsection{Can AFD-Ledger Be Trusted?}
\label{sec:validation_overview}

Before using AFD-Ledger to study deployment behavior, we validate its two key design components: analytical deployment provisioning and bounded hardware exploration.

\subsubsection{Deployment Provisioning Validation}
\label{sec:validation}

AFD-Ledger evaluates candidate deployments using the analytical execution model described in Section~\ref{sec:background}. Because the model omits runtime-specific overheads such as kernel deficiency, scheduler latency and network contention, we compare its provisioning decisions with measurements from our LongCat~2.0 runtime under the same deployment specification. The runtime uses the best deployment supported by our implementation (128 attention cards and 32 FFN cards), while AFD-Ledger independently reprovisions both deployment families under the same hardware budget and workload.

\begin{figure}[t]
\centering
\includegraphics[width=\linewidth]{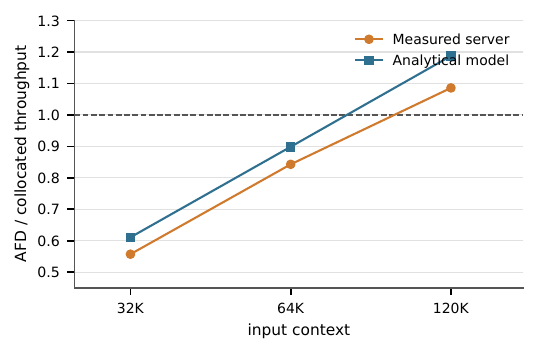}
\caption{Measured and analytical AFD-to-collocated throughput ratios on three LongCat~2.0 validation workloads. The dashed line denotes architecture parity.}
\label{fig:validation}
\end{figure}

Figure~\ref{fig:validation} shows that analytical provisioning and the
physical runtime make the same deployment decision on all three workloads:
collocated serving is preferred at short contexts, while AFD becomes
preferable at longer contexts. Although the analytical model overestimates the
absolute AFD gain by omitting runtime overheads, the throughput ratios remain
within 6.6--9.6\% of measurement, preserving the deployment ordering required
for provisioning.

\subsubsection{Hardware Exploration Validation}
\label{sec:search_validation}

We next evaluate whether the bounded hardware exploration introduced in Section~\ref{sec:hardware_search} can identify the same deployment as exhaustive hardware evaluation while requiring substantially fewer complete deployment provisions.

\paragraph{Recovery of Exhaustive Provisioning}

We next evaluate whether bounded hardware exploration recovers exhaustive provisioning while evaluating substantially fewer hardware assignments.

\begin{figure}[t]
\centering
\includegraphics[width=\linewidth]{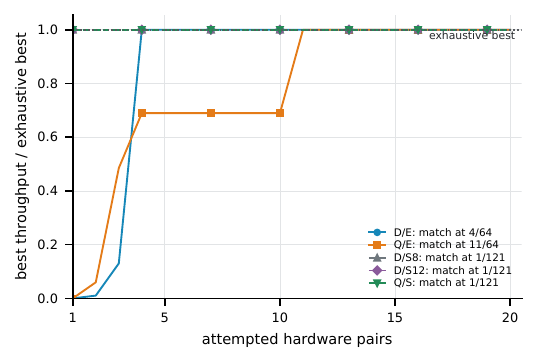}
\caption{Best-so-far modeled AFD throughput over 20 attempted pairs for the
five reference cases in Table~\ref{tab:reference_cases}, normalized to
exhaustive ordered-pair enumeration.}
\label{fig:convergence}
\end{figure}

Figure~\ref{fig:convergence} shows that AFD-Ledger reaches the exhaustive
optimum within the 20-assignment budget for all five reference cases. This
reduces complete deployment evaluations from 64 to 20 assignments for Catalog
E and from 121 to 20 for Catalog S, eliminating 68.8\% and 83.5\% of
provisioning runs, respectively. Whenever exhaustive search is feasible, we
report the remaining throughput gap to quantify search quality directly.

\paragraph{Exploration Strategy Ablation}

We next evaluate whether the bounded exploration strategy can be simplified without sacrificing deployment quality.

\begin{figure}[t]
\centering
\includegraphics[width=\linewidth]{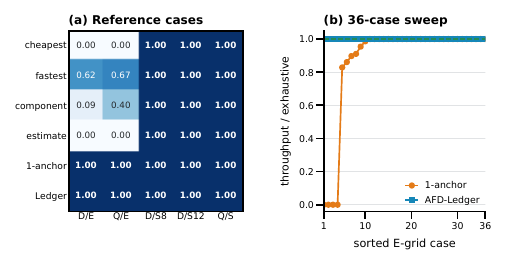}
\caption{Search-policy comparison normalized to exhaustive AFD throughput.
(a) The five reference cases in Table~\ref{tab:reference_cases}.
(b) Sorted results over the 36 fully enumerated E-grid points.
Labels denote minimum pair price (cheapest), maximum raw role service
(fastest), component-wise memory/compute per price (component), maximum
modeled role service per price (estimate only), single-seed local refinement
(one anchor), and role-diverse bounded refinement (AFD-Ledger).
The one-anchor ablation uses the same expansion width and round limit but no
multi-anchor start, role-diverse beam, or final globally ranked pass.}
\label{fig:search_ablation}
\end{figure}

Figure~\ref{fig:search_ablation}(a) compares AFD-Ledger with several simpler exploration policies, including static hardware ranking and single-anchor local refinement. Static hardware rankings often select infeasible or suboptimal hardware assignments because they ignore deployment organization. Single-anchor refinement performs well on the five reference cases but misses the global optimum on more diverse deployment specifications.

Figure~\ref{fig:search_ablation}(b) extends the comparison to the 36 fully enumerated E-grid deployments. Under the same evaluation budget, the one-anchor search misses the optimum in ten cases, including four infeasible selections, whereas AFD-Ledger consistently recovers the exhaustive optimum. These results show that combining multiple starting anchors, role-diverse retention, and the final global exploration is necessary for robust bounded hardware exploration.

\subsection{What Deployment Decisions Does AFD-Ledger Reveal?}
\label{sec:deployment_insights}

Having validated AFD-Ledger, we now use it to study deployment behavior across models, hardware catalogs, and deployment specifications. Three consistent deployment insights emerge.

\subsubsection{Homogeneous AFD Provides Higher Throughput in Only a Minority of Studied Settings}
\label{sec:existing_results}

We first ask whether simply separating attention and expert execution is sufficient to improve serving throughput. To isolate the effect of disaggregation itself, we restrict AFD to homogeneous deployments, in which attention and expert workers use the same hardware, and independently provision both AFD and collocated deployments across the complete E-grid.

Across the 36 deployment specifications in E-grid, AFD-Ledger selects homogeneous AFD in only seven cases: none of the 18 DeepSeek configurations and seven of the 18 Qwen configurations (Figure~\ref{fig:existing_results}(a)). In all remaining settings, the optimized collocated deployment achieves higher throughput.

The fixed-budget ledger explains this behavior. Separating experts increases batch size and can reduce TPOT, but it also removes request-serving GPUs. In most deployment specifications, the batch and latency gains are insufficient to offset the resulting loss in request capacity.

The few successful cases illustrate when homogeneous AFD can still be beneficial. The strongest example is Qwen on A100 PCIe under a budget of 64 and a 20-ms TPOT target. Compared with the optimized collocated deployment, AFD reduces the worker factor from \(145/192=0.76\), but doubles the resident batch from 32 to 64 while leaving TPOT nearly unchanged. The larger batch factor therefore outweighs the worker loss, resulting in a final throughput ratio of \(1.51\times\).

\subsubsection{Heterogeneous AFD Requires the Right Hardware Complementarity}

We next allow attention and expert execution to use different hardware.

\begin{figure}[t]
\centering
\includegraphics[width=\linewidth]{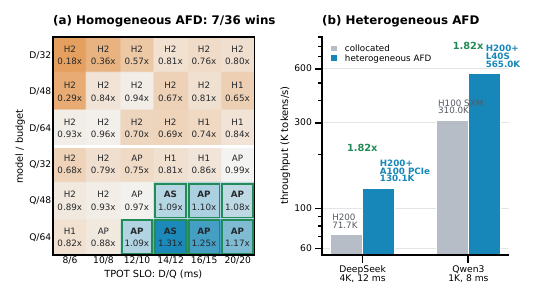}
\caption{Existing-catalog outcomes. D=DeepSeek, Q=Qwen3; H2, H1, AP, and AS denote H200, H100 SXM, A100 PCIe, and
A100 SXM.}
\label{fig:existing_results}
\end{figure}

Figure~\ref{fig:existing_results}(b) shows two representative deployment specifications where AFD-Ledger selects heterogeneous AFD. For DeepSeek, the best deployment combines H200 attention workers with A100 PCIe expert workers, achieving \(1.815\times\) the throughput of the best collocated deployment. For Qwen, the selected deployment combines H200 attention workers with L40S expert workers and reaches \(1.823\times\) the collocated throughput.

Using lower-cost hardware for expert execution frees more of the fixed budget for attention workers, increasing resident concurrency while keeping TPOT close to the collocated deployment. The resulting throughput gain therefore comes primarily from higher concurrency rather than lower latency.

\begin{figure}[t]
\centering
\includegraphics[width=\linewidth]{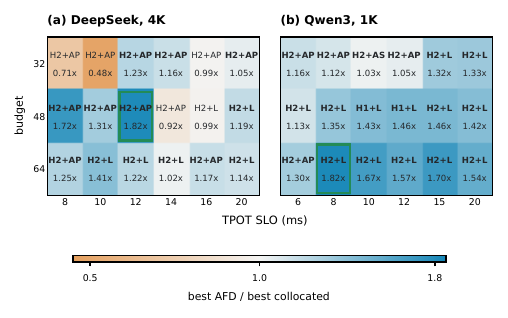}
\caption{Fully provisioned E-grid decisions for DeepSeek (a) and Qwen (b). H2 and H1 denote H200 and H100; AP and AS denote A100 PCIe and A100 SXM;
L denotes L40S.}
\label{fig:existing_regions}
\end{figure}

Figure~\ref{fig:existing_regions} further shows that no hardware pair is
universally optimal: the same H200+A100 PCIe combination ranges from
0.483$\times$ to 1.815$\times$ depending on the deployment specification. Changing the budget or TPOT target changes the optimal deployment
organization, including worker allocation, batching, and even the collocated
reference. Hardware complementarity therefore emerges only after complete
deployment provisioning.

\subsubsection{Role-Specific Hardware Improvements Matter Most When They Unlock Better Deployment Organizations}

Finally, we study how future role-specialized hardware changes deployment behavior.

\begin{figure}[t]
\centering
\includegraphics[width=\linewidth]{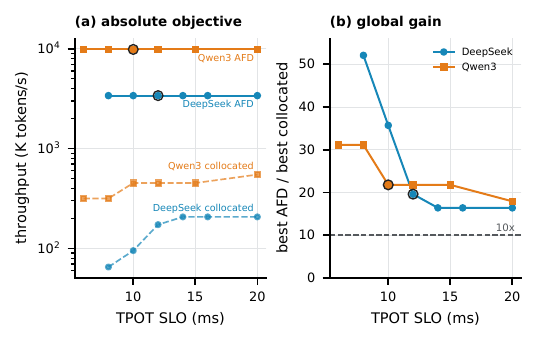}
\caption{Provisioning with the fixed S catalog. Black rings mark D/S12 and Q/S from
Table~\ref{tab:reference_cases}.}
\label{fig:simulated_slo}
\end{figure}

Figure~\ref{fig:simulated_slo} shows that the simulated hardware consistently changes the deployment decision even when the strongest commercial Blackwell devices remain available to collocated serving. Across the studied S-sweep, AFD achieves throughput improvements ranging from \(16.39\times\) to \(52.07\times\) for DeepSeek and from \(17.92\times\) to \(31.12\times\) for Qwen.

Unlike the previous heterogeneous deployments, the dominant gain now comes
from enabling a different deployment organization rather than improving one
component in isolation. In the representative D/S12 case, the lower per-replica
cost increases the number of replicas from one to three, raising supported
concurrency from 1,856 to 20,544 requests and yielding a 19.59$\times$
throughput improvement.

More broadly, hardware improvements matter only when they unlock a better
deployment organization, such as additional replicas, larger resident batches,
or different worker ratios. Evaluating future AFD hardware therefore requires
measuring the deployments it enables rather than isolated hardware metrics.

\subsection{How Robust Are the Deployment Decisions?}
\label{sec:sensitivity}

We finally evaluate whether the deployment decisions remain stable under
changes to hardware capabilities and analytical assumptions. In each
experiment, AFD-Ledger reprovisions the deployment after modifying one
parameter.

\subsubsection{Hardware Sensitivity}

Starting from the simulated S catalog, we vary one hardware property at a time
and rerun complete provisioning.

\begin{figure}[t]
\centering
\includegraphics[width=\linewidth]{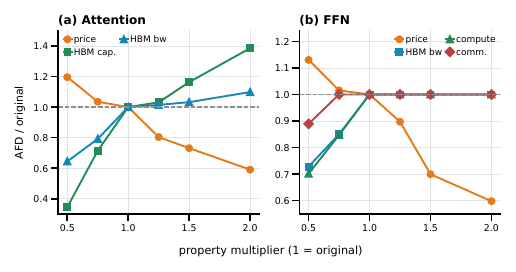}
\caption{One-at-a-time hardware sensitivity for the representative D/S12
deployment. Throughput is normalized to the original
S-catalog result.}
\label{fig:hardware_sensitivity}
\end{figure}

Figure~\ref{fig:hardware_sensitivity} shows that reducing device cost or
increasing memory capacity has the largest impact because these changes enable
better deployment organizations. In contrast, increasing compute throughput or
memory bandwidth alone often yields limited benefit once another stage becomes
the bottleneck. The largest gains therefore come from reprovisioning rather than accelerating
a fixed deployment, reinforcing the conclusions of
Section~\ref{sec:deployment_insights}.

\subsubsection{Runtime Overhead Robustness}

We next evaluate whether moderate runtime overhead changes the deployment
decisions.

\begin{figure}[t]
\centering
\includegraphics[width=\linewidth]{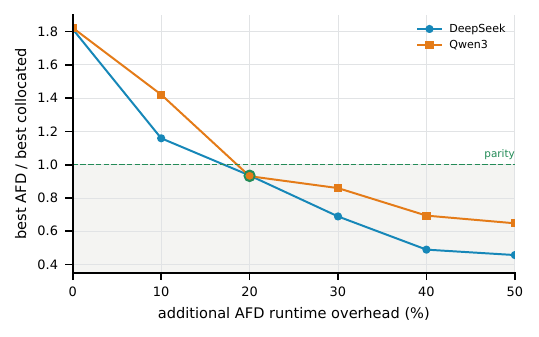}
\caption{Deployment decisions under additional AFD runtime overhead for the
D/E and Q/E reference cases.}
\label{fig:overhead_sensitivity}
\end{figure}

Figure~\ref{fig:overhead_sensitivity} shows that the selected AFD deployments
remain preferable under approximately 10\% additional runtime overhead.
Around 20\%, both cases switch to collocated serving because reprovisioning
selects different deployment organizations after some AFD candidates become
TPOT-infeasible. These results indicate that AFD-Ledger is robust to moderate modeling error and
that runtime overhead primarily affects deployment quality through
reprovisioning.

Overall, both studies show that deployment decisions are more sensitive to the
deployment organizations enabled by hardware and software changes than to the
raw improvements of individual device metrics.

\section{Discussion}
\label{sec:discussion}

Although motivated by AFD, AFD-Ledger is applicable to any deployment architecture that can be provisioned under a common deployment specification and evaluated with a shared throughput objective. Supporting a new architecture requires only an architecture-specific deployment enumerator and performance model, while the hardware exploration and comparison framework remain unchanged. More broadly, this work suggests that deployment provisioning can serve not only as an optimization tool but also as a principled methodology for comparing system architectures under consistent deployment constraints.

Our conclusions are conditioned on the fidelity of the analytical performance model and the evaluated hardware catalogs. The current implementation intentionally omits implementation-specific effects such as software overheads and runtime contention to keep provisioning efficient. Although physical validation shows that these simplifications preserve deployment decisions in our experiments, higher-fidelity performance models could further improve accuracy. Likewise, our future-hardware study evaluates hypothetical capability points rather than predicting commercial products.

Finally, the bounded hardware exploration is heuristic rather than theoretically optimal. While it consistently matches exhaustive provisioning in our evaluation, larger hardware catalogs may require more sophisticated search strategies. Nevertheless, we believe AFD-Ledger demonstrates that deployment architectures should be evaluated through independently optimized deployments under a common provisioning framework rather than manually selected configurations.

\section{Related Work}
\label{sec:related}

Prior work on LLM serving primarily optimizes execution within a deployment through batching, scheduling, KV-cache management, parallelism, communication, and kernel optimizations~\cite{yu2022orca,Sarathi2023,SarathiServe2024,liu2025lmcache,liu2026kvserve,guo2026splitzip,DeepSpeedMoE2022,FastMoE2021,Tutel2022,SGLangDeepSeekEP2025,SGLangAntGroup2025}. Disaggregated serving extends these techniques by separating execution roles, including prefill--decode disaggregation, KV-centric architectures, memory--compute disaggregation, and AFD~\cite{Splitwise2023,DistServe2024,hu2024inference,qin2024mooncake,hu2024memserve,ModelAttentionDisagg2025,MegaScaleInfer2025,Step32025,xiao2025xdeepserve,zhang2025janus,pan2025efficient,EaaS2025,fastafd2026,RevealingAFD2026}. Existing AFD systems focus on runtime mechanisms such as scheduling, communication, elasticity, and hardware placement within AFD deployments. More recently, simulator-based studies have shown that deployment behavior can be analyzed without implementing every candidate system~\cite{AnalyticalAFD2026,HowFarDisaggregation2026,Frontier2026}, but they either study fixed hardware pools or optimize within predetermined deployment spaces. In contrast, AFD-Ledger treats provisioning itself as the problem, jointly exploring hardware assignment and deployment organization under a fixed-budget deployment contract while bounding the number of complete deployment evaluations.

\section{Conclusion}

This paper presented \emph{AFD-Ledger}, a provisioning system for jointly exploring hardware assignment and deployment organization when comparing collocated serving and AFD. By combining analytical deployment provisioning with bounded hardware exploration, AFD-Ledger efficiently searches complete deployments while avoiding exhaustive hardware evaluation. Our results show that the benefits of AFD depend on deployment-level provisioning rather than isolated hardware or component metrics, and that complete provisioning can reveal both practical deployment decisions and hardware-design opportunities. We hope AFD-Ledger provides a useful foundation for future deployment provisioning systems and deployment-aware hardware--software co-design.

\bibliography{ref}


\begin{thebibliography}{36}


\ifx \showCODEN    \undefined \def \showCODEN     #1{\unskip}     \fi
\ifx \showISBNx    \undefined \def \showISBNx     #1{\unskip}     \fi
\ifx \showISBNxiii \undefined \def \showISBNxiii  #1{\unskip}     \fi
\ifx \showISSN     \undefined \def \showISSN      #1{\unskip}     \fi
\ifx \showLCCN     \undefined \def \showLCCN      #1{\unskip}     \fi
\ifx \shownote     \undefined \def \shownote      #1{#1}          \fi
\ifx \showarticletitle \undefined \def \showarticletitle #1{#1}   \fi
\ifx \showURL      \undefined \def \showURL       {\relax}        \fi
\providecommand\bibfield[2]{#2}
\providecommand\bibinfo[2]{#2}
\providecommand\natexlab[1]{#1}
\providecommand\showeprint[2][]{arXiv:#2}

\bibitem[Run(2026)]%
        {RunPodPricing2026}
 \bibinfo{year}{2026}\natexlab{}.
\newblock \bibinfo{title}{{RunPod} GPU Cloud Pricing}.
\newblock \bibinfo{howpublished}{\url{https://www.runpod.io/pricing}}.
\newblock
\newblock
\shownote{Accessed July 5, 2026}.


\bibitem[Agrawal et~al\mbox{.}(2024)]%
        {SarathiServe2024}
\bibfield{author}{\bibinfo{person}{Amey Agrawal}, \bibinfo{person}{Nitin
  Kedia}, \bibinfo{person}{Ashish Panwar}, \bibinfo{person}{Jayashree Mohan},
  \bibinfo{person}{Nipun Kwatra}, \bibinfo{person}{Bhargav Gulavani},
  \bibinfo{person}{Alexey Tumanov}, {and} \bibinfo{person}{Ramachandran
  Ramjee}.} \bibinfo{year}{2024}\natexlab{}.
\newblock \showarticletitle{Taming $\{$Throughput-Latency$\}$ tradeoff in
  $\{$LLM$\}$ inference with $\{$Sarathi-Serve$\}$}. In
  \bibinfo{booktitle}{\emph{18th USENIX symposium on operating systems design
  and implementation (OSDI 24)}}. \bibinfo{pages}{117--134}.
\newblock


\bibitem[Agrawal et~al\mbox{.}(2023)]%
        {Sarathi2023}
\bibfield{author}{\bibinfo{person}{Amey Agrawal}, \bibinfo{person}{Ashish
  Panwar}, \bibinfo{person}{Jayashree Mohan}, \bibinfo{person}{Nipun Kwatra},
  \bibinfo{person}{Bhargav~S Gulavani}, {and} \bibinfo{person}{Ramachandran
  Ramjee}.} \bibinfo{year}{2023}\natexlab{}.
\newblock \showarticletitle{Sarathi: Efficient llm inference by piggybacking
  decodes with chunked prefills}.
\newblock \bibinfo{journal}{\emph{arXiv preprint arXiv:2308.16369}}
  (\bibinfo{year}{2023}).
\newblock


\bibitem[Chen et~al\mbox{.}(2024)]%
        {ModelAttentionDisagg2025}
\bibfield{author}{\bibinfo{person}{Shaoyuan Chen}, \bibinfo{person}{Wencong
  Xiao}, \bibinfo{person}{Yutong Lin}, \bibinfo{person}{Mingxing Zhang},
  \bibinfo{person}{Yingdi Shan}, \bibinfo{person}{Jinlei Jiang},
  \bibinfo{person}{Kang Chen}, {and} \bibinfo{person}{Yongwei Wu}.}
  \bibinfo{year}{2024}\natexlab{}.
\newblock \showarticletitle{Efficient heterogeneous large language model
  decoding with model-attention disaggregation}.
\newblock \bibinfo{journal}{\emph{arXiv preprint arXiv:2405.01814}}
  (\bibinfo{year}{2024}).
\newblock


\bibitem[Fedus et~al\mbox{.}(2022)]%
        {SwitchTransformer2021}
\bibfield{author}{\bibinfo{person}{William Fedus}, \bibinfo{person}{Barret
  Zoph}, {and} \bibinfo{person}{Noam Shazeer}.}
  \bibinfo{year}{2022}\natexlab{}.
\newblock \showarticletitle{Switch transformers: Scaling to trillion parameter
  models with simple and efficient sparsity}.
\newblock \bibinfo{journal}{\emph{Journal of Machine Learning Research}}
  \bibinfo{volume}{23}, \bibinfo{number}{120} (\bibinfo{year}{2022}),
  \bibinfo{pages}{1--39}.
\newblock


\bibitem[Feng et~al\mbox{.}(2026)]%
        {Frontier2026}
\bibfield{author}{\bibinfo{person}{Yicheng Feng}, \bibinfo{person}{Xin Tan},
  \bibinfo{person}{Yangtao Deng}, \bibinfo{person}{Yimin Jiang},
  \bibinfo{person}{Yibo Zhu}, {and} \bibinfo{person}{Hong Xu}.}
  \bibinfo{year}{2026}\natexlab{}.
\newblock \showarticletitle{Frontier: Towards Comprehensive and Accurate LLM
  Inference Simulation}.
\newblock \bibinfo{journal}{\emph{arXiv preprint arXiv:2605.21312}}
  (\bibinfo{year}{2026}).
\newblock


\bibitem[Fu et~al\mbox{.}(2026)]%
        {fastafd2026}
\bibfield{author}{\bibinfo{person}{Yichao Fu}, \bibinfo{person}{Yuxuan Zhang},
  \bibinfo{person}{Ruitian Wang}, \bibinfo{person}{Junda Chen}, {and}
  \bibinfo{person}{Hao Zhang}.} \bibinfo{year}{2026}\natexlab{}.
\newblock \bibinfo{title}{FastAFD: Open-Source Large-Scale Attention-FFN
  Disaggregation on Blackwell NVL72}.
\newblock
\urldef\tempurl%
\url{https://github.com/hao-ai-lab/FastAFD}
\showURL{%
\tempurl}
\newblock
\shownote{Technical blog and open-source release}.


\bibitem[Guo and Joshi(2026)]%
        {guo2026splitzip}
\bibfield{author}{\bibinfo{person}{Yipin Guo} {and} \bibinfo{person}{Siddharth
  Joshi}.} \bibinfo{year}{2026}\natexlab{}.
\newblock \showarticletitle{SplitZip: Ultra Fast Lossless KV Compression for
  Disaggregated LLM Serving}.
\newblock \bibinfo{journal}{\emph{arXiv preprint arXiv:2605.01708}}
  (\bibinfo{year}{2026}).
\newblock


\bibitem[He et~al\mbox{.}(2021)]%
        {FastMoE2021}
\bibfield{author}{\bibinfo{person}{Jiaao He}, \bibinfo{person}{Jiezhong Qiu},
  \bibinfo{person}{Aohan Zeng}, \bibinfo{person}{Zhilin Yang},
  \bibinfo{person}{Jidong Zhai}, {and} \bibinfo{person}{Jie Tang}.}
  \bibinfo{year}{2021}\natexlab{}.
\newblock \showarticletitle{Fastmoe: A fast mixture-of-expert training system}.
\newblock \bibinfo{journal}{\emph{arXiv preprint arXiv:2103.13262}}
  (\bibinfo{year}{2021}).
\newblock


\bibitem[Hu et~al\mbox{.}(2024a)]%
        {hu2024memserve}
\bibfield{author}{\bibinfo{person}{Cunchen Hu}, \bibinfo{person}{Heyang Huang},
  \bibinfo{person}{Junhao Hu}, \bibinfo{person}{Jiang Xu},
  \bibinfo{person}{Xusheng Chen}, \bibinfo{person}{Tao Xie},
  \bibinfo{person}{Chenxi Wang}, \bibinfo{person}{Sa Wang},
  \bibinfo{person}{Yungang Bao}, \bibinfo{person}{Ninghui Sun},
  {et~al\mbox{.}}} \bibinfo{year}{2024}\natexlab{a}.
\newblock \showarticletitle{Memserve: Context caching for disaggregated llm
  serving with elastic memory pool}.
\newblock \bibinfo{journal}{\emph{arXiv preprint arXiv:2406.17565}}
  (\bibinfo{year}{2024}).
\newblock


\bibitem[Hu et~al\mbox{.}(2024b)]%
        {hu2024inference}
\bibfield{author}{\bibinfo{person}{Cunchen Hu}, \bibinfo{person}{Heyang Huang},
  \bibinfo{person}{Liangliang Xu}, \bibinfo{person}{Xusheng Chen},
  \bibinfo{person}{Jiang Xu}, \bibinfo{person}{Shuang Chen},
  \bibinfo{person}{Hao Feng}, \bibinfo{person}{Chenxi Wang},
  \bibinfo{person}{Sa Wang}, \bibinfo{person}{Yungang Bao}, {et~al\mbox{.}}}
  \bibinfo{year}{2024}\natexlab{b}.
\newblock \showarticletitle{Inference without interference: Disaggregate llm
  inference for mixed downstream workloads}.
\newblock \bibinfo{journal}{\emph{arXiv preprint arXiv:2401.11181}}
  (\bibinfo{year}{2024}).
\newblock


\bibitem[Hwang et~al\mbox{.}(2023)]%
        {Tutel2022}
\bibfield{author}{\bibinfo{person}{Changho Hwang}, \bibinfo{person}{Wei Cui},
  \bibinfo{person}{Yifan Xiong}, \bibinfo{person}{Ziyue Yang},
  \bibinfo{person}{Ze Liu}, \bibinfo{person}{Han Hu}, \bibinfo{person}{Zilong
  Wang}, \bibinfo{person}{Rafael Salas}, \bibinfo{person}{Jithin Jose},
  \bibinfo{person}{Prabhat Ram}, {et~al\mbox{.}}}
  \bibinfo{year}{2023}\natexlab{}.
\newblock \showarticletitle{Tutel: Adaptive mixture-of-experts at scale}.
\newblock \bibinfo{journal}{\emph{Proceedings of Machine Learning and Systems}}
   \bibinfo{volume}{5} (\bibinfo{year}{2023}), \bibinfo{pages}{269--287}.
\newblock


\bibitem[Lepikhin et~al\mbox{.}(2020)]%
        {GShard2020}
\bibfield{author}{\bibinfo{person}{Dmitry Lepikhin},
  \bibinfo{person}{HyoukJoong Lee}, \bibinfo{person}{Yuanzhong Xu},
  \bibinfo{person}{Dehao Chen}, \bibinfo{person}{Orhan Firat},
  \bibinfo{person}{Yanping Huang}, \bibinfo{person}{Maxim Krikun},
  \bibinfo{person}{Noam Shazeer}, {and} \bibinfo{person}{Zhifeng Chen}.}
  \bibinfo{year}{2020}\natexlab{}.
\newblock \showarticletitle{Gshard: Scaling giant models with conditional
  computation and automatic sharding}.
\newblock \bibinfo{journal}{\emph{arXiv preprint arXiv:2006.16668}}
  (\bibinfo{year}{2020}).
\newblock


\bibitem[Liu et~al\mbox{.}(2025b)]%
        {DeepSeekV32}
\bibfield{author}{\bibinfo{person}{Aixin Liu}, \bibinfo{person}{Aoxue Mei},
  \bibinfo{person}{Bangcai Lin}, \bibinfo{person}{Bing Xue},
  \bibinfo{person}{Bingxuan Wang}, \bibinfo{person}{Bingzheng Xu},
  \bibinfo{person}{Bochao Wu}, \bibinfo{person}{Bowei Zhang},
  \bibinfo{person}{Chaofan Lin}, \bibinfo{person}{Chen Dong}, {et~al\mbox{.}}}
  \bibinfo{year}{2025}\natexlab{b}.
\newblock \showarticletitle{Deepseek-v3. 2: Pushing the frontier of open large
  language models}.
\newblock \bibinfo{journal}{\emph{arXiv preprint arXiv:2512.02556}}
  (\bibinfo{year}{2025}).
\newblock


\bibitem[Liu et~al\mbox{.}(2026a)]%
        {RevealingAFD2026}
\bibfield{author}{\bibinfo{person}{Guowei Liu}, \bibinfo{person}{Hongming Li},
  \bibinfo{person}{Yaning Guo}, \bibinfo{person}{Yongxi Lyu},
  \bibinfo{person}{Mo Zhou}, \bibinfo{person}{Yi Liu},
  \bibinfo{person}{Zhaogeng Li}, {and} \bibinfo{person}{Yanpeng Wang}.}
  \bibinfo{year}{2026}\natexlab{a}.
\newblock \showarticletitle{Revealing the Challenges of Attention-FFN
  Disaggregation for Modern MoE Models and Hardware Systems}.
\newblock \bibinfo{journal}{\emph{arXiv preprint arXiv:2602.09721}}
  (\bibinfo{year}{2026}).
\newblock


\bibitem[Liu et~al\mbox{.}(2025a)]%
        {liu2025lmcache}
\bibfield{author}{\bibinfo{person}{Yuhan Liu}, \bibinfo{person}{Yihua Cheng},
  \bibinfo{person}{Jiayi Yao}, \bibinfo{person}{Yuwei An},
  \bibinfo{person}{Xiaokun Chen}, \bibinfo{person}{Shaoting Feng},
  \bibinfo{person}{Yuyang Huang}, \bibinfo{person}{Samuel Shen},
  \bibinfo{person}{Rui Zhang}, \bibinfo{person}{Kuntai Du}, {et~al\mbox{.}}}
  \bibinfo{year}{2025}\natexlab{a}.
\newblock \showarticletitle{Lmcache: An efficient KV cache layer for
  enterprise-scale LLM inference}.
\newblock \bibinfo{journal}{\emph{arXiv preprint arXiv:2510.09665}}
  (\bibinfo{year}{2025}).
\newblock


\bibitem[Liu et~al\mbox{.}(2026b)]%
        {liu2026kvserve}
\bibfield{author}{\bibinfo{person}{Zedong Liu}, \bibinfo{person}{Xinyang Ma},
  \bibinfo{person}{Dejun Luo}, \bibinfo{person}{Hairui Zhao},
  \bibinfo{person}{Bing Lu}, \bibinfo{person}{Wenjing Huang},
  \bibinfo{person}{Yida Gu}, \bibinfo{person}{Xingchen Liu},
  \bibinfo{person}{Zheng Wei}, \bibinfo{person}{Jinyang Liu}, {et~al\mbox{.}}}
  \bibinfo{year}{2026}\natexlab{b}.
\newblock \showarticletitle{KVServe: Service-Aware KV Cache Compression for
  Communication-Efficient Disaggregated LLM Serving}.
\newblock \bibinfo{journal}{\emph{arXiv preprint arXiv:2605.13734}}
  (\bibinfo{year}{2026}).
\newblock


\bibitem[Liu et~al\mbox{.}(2025c)]%
        {EaaS2025}
\bibfield{author}{\bibinfo{person}{Ziming Liu}, \bibinfo{person}{Boyu Tian},
  \bibinfo{person}{Guoteng Wang}, \bibinfo{person}{Zhen Jiang},
  \bibinfo{person}{Peng Sun}, \bibinfo{person}{Zhenhua Han},
  \bibinfo{person}{Tian Tang}, \bibinfo{person}{Xiaohe Hu},
  \bibinfo{person}{Yanmin Jia}, \bibinfo{person}{Yan Zhang}, {et~al\mbox{.}}}
  \bibinfo{year}{2025}\natexlab{c}.
\newblock \showarticletitle{Expert-as-a-service: Towards efficient, scalable,
  and robust large-scale moe serving}.
\newblock \bibinfo{journal}{\emph{arXiv preprint arXiv:2509.17863}}
  (\bibinfo{year}{2025}).
\newblock


\bibitem[{Meituan LongCat Team}(2026)]%
        {LongCat20Model2026}
\bibfield{author}{\bibinfo{person}{{Meituan LongCat Team}}.}
  \bibinfo{year}{2026}\natexlab{}.
\newblock \bibinfo{title}{{LongCat-2.0}}.
\newblock
  \bibinfo{howpublished}{\url{https://huggingface.co/meituan-longcat/LongCat-2.0}}.
\newblock
\newblock
\shownote{Accessed August 1, 2026}.


\bibitem[{NVIDIA Corporation}(2026)]%
        {NVIDIADataCenterGPUs2026}
\bibfield{author}{\bibinfo{person}{{NVIDIA Corporation}}.}
  \bibinfo{year}{2026}\natexlab{}.
\newblock \bibinfo{title}{{NVIDIA} Data Center {GPU} Product Specifications}.
\newblock
  \bibinfo{howpublished}{\url{https://www.nvidia.com/en-us/data-center/data-center-gpus/}}.
\newblock
\newblock
\shownote{Accessed July 10, 2026}.


\bibitem[Pan et~al\mbox{.}(2025)]%
        {pan2025efficient}
\bibfield{author}{\bibinfo{person}{Xinglin Pan}, \bibinfo{person}{Shaohuai
  Shi}, \bibinfo{person}{Wenxiang Lin}, \bibinfo{person}{Yuxin Wang},
  \bibinfo{person}{Zhenheng Tang}, \bibinfo{person}{Wei Wang}, {and}
  \bibinfo{person}{Xiaowen Chu}.} \bibinfo{year}{2025}\natexlab{}.
\newblock \showarticletitle{Efficient MoE Inference with Fine-Grained
  Scheduling of Disaggregated Expert Parallelism}.
\newblock \bibinfo{journal}{\emph{arXiv preprint arXiv:2512.21487}}
  (\bibinfo{year}{2025}).
\newblock


\bibitem[Patel et~al\mbox{.}(2024)]%
        {Splitwise2023}
\bibfield{author}{\bibinfo{person}{Pratyush Patel}, \bibinfo{person}{Esha
  Choukse}, \bibinfo{person}{Chaojie Zhang}, \bibinfo{person}{Aashaka Shah},
  \bibinfo{person}{{\'I}{\~n}igo Goiri}, \bibinfo{person}{Saeed Maleki}, {and}
  \bibinfo{person}{Ricardo Bianchini}.} \bibinfo{year}{2024}\natexlab{}.
\newblock \showarticletitle{Splitwise: Efficient generative llm inference using
  phase splitting}. In \bibinfo{booktitle}{\emph{2024 ACM/IEEE 51st Annual
  International Symposium on Computer Architecture (ISCA)}}. IEEE,
  \bibinfo{pages}{118--132}.
\newblock


\bibitem[Qin et~al\mbox{.}(2024)]%
        {qin2024mooncake}
\bibfield{author}{\bibinfo{person}{Ruoyu Qin}, \bibinfo{person}{Zheming Li},
  \bibinfo{person}{Weiran He}, \bibinfo{person}{Jialei Cui},
  \bibinfo{person}{Heyi Tang}, \bibinfo{person}{Feng Ren},
  \bibinfo{person}{Teng Ma}, \bibinfo{person}{Shangming Cai},
  \bibinfo{person}{Yineng Zhang}, \bibinfo{person}{Mingxing Zhang},
  {et~al\mbox{.}}} \bibinfo{year}{2024}\natexlab{}.
\newblock \showarticletitle{Mooncake: A kvcache-centric disaggregated
  architecture for llm serving}.
\newblock \bibinfo{journal}{\emph{ACM Transactions on Storage}}
  (\bibinfo{year}{2024}).
\newblock


\bibitem[Rajbhandari et~al\mbox{.}(2022)]%
        {DeepSpeedMoE2022}
\bibfield{author}{\bibinfo{person}{Samyam Rajbhandari},
  \bibinfo{person}{Conglong Li}, \bibinfo{person}{Zhewei Yao},
  \bibinfo{person}{Minjia Zhang}, \bibinfo{person}{Reza~Yazdani Aminabadi},
  \bibinfo{person}{Ammar~Ahmad Awan}, \bibinfo{person}{Jeff Rasley}, {and}
  \bibinfo{person}{Yuxiong He}.} \bibinfo{year}{2022}\natexlab{}.
\newblock \showarticletitle{Deepspeed-moe: Advancing mixture-of-experts
  inference and training to power next-generation ai scale}. In
  \bibinfo{booktitle}{\emph{International conference on machine learning}}.
  PMLR, \bibinfo{pages}{18332--18346}.
\newblock


\bibitem[Song et~al\mbox{.}(2026)]%
        {AnalyticalAFD2026}
\bibfield{author}{\bibinfo{person}{Chendong Song}, \bibinfo{person}{Meixuan
  Wang}, \bibinfo{person}{Hang Zhou}, \bibinfo{person}{Hong Liang},
  \bibinfo{person}{Yuan Lyu}, \bibinfo{person}{Zixi Chen},
  \bibinfo{person}{Yuwei Fan}, {and} \bibinfo{person}{Zijie Zhou}.}
  \bibinfo{year}{2026}\natexlab{}.
\newblock \showarticletitle{Analytical Provisioning for Attention-FFN
  Disaggregated LLM Serving under Stochastic Workloads}.
\newblock \bibinfo{journal}{\emph{arXiv preprint arXiv:2601.21351}}
  (\bibinfo{year}{2026}).
\newblock


\bibitem[Team et~al\mbox{.}(2025)]%
        {LongCatFlash2025}
\bibfield{author}{\bibinfo{person}{Meituan~LongCat Team}, \bibinfo{person}{Bei
  Li}, \bibinfo{person}{Bingye Lei}, \bibinfo{person}{Bo Wang},
  \bibinfo{person}{Bolin Rong}, \bibinfo{person}{Chao Wang},
  \bibinfo{person}{Chao Zhang}, \bibinfo{person}{Chen Gao},
  \bibinfo{person}{Chen Zhang}, \bibinfo{person}{Cheng Sun}, {et~al\mbox{.}}}
  \bibinfo{year}{2025}\natexlab{}.
\newblock \showarticletitle{Longcat-flash technical report}.
\newblock \bibinfo{journal}{\emph{arXiv preprint arXiv:2509.01322}}
  (\bibinfo{year}{2025}).
\newblock


\bibitem[{The SGLang Team}(2025)]%
        {SGLangDeepSeekEP2025}
\bibfield{author}{\bibinfo{person}{{The SGLang Team}}.}
  \bibinfo{year}{2025}\natexlab{}.
\newblock \bibinfo{title}{Deploying {DeepSeek} with {PD} Disaggregation and
  Large-Scale Expert Parallelism on 96 {H100} {GPUs}}.
\newblock
  \bibinfo{howpublished}{\url{https://lmsys.org/blog/2025-05-05-large-scale-ep}}.
\newblock


\bibitem[Wang et~al\mbox{.}(2025)]%
        {Step32025}
\bibfield{author}{\bibinfo{person}{Bin Wang}, \bibinfo{person}{Bojun Wang},
  \bibinfo{person}{Changyi Wan}, \bibinfo{person}{Guanzhe Huang},
  \bibinfo{person}{Hanpeng Hu}, \bibinfo{person}{Haonan Jia},
  \bibinfo{person}{Hao Nie}, \bibinfo{person}{Mingliang Li},
  \bibinfo{person}{Nuo Chen}, \bibinfo{person}{Siyu Chen}, {et~al\mbox{.}}}
  \bibinfo{year}{2025}\natexlab{}.
\newblock \showarticletitle{Step-3 is large yet affordable: Model-system
  co-design for cost-effective decoding}.
\newblock \bibinfo{journal}{\emph{arXiv preprint arXiv:2507.19427}}
  (\bibinfo{year}{2025}).
\newblock


\bibitem[Wu et~al\mbox{.}(2026)]%
        {HowFarDisaggregation2026}
\bibfield{author}{\bibinfo{person}{Hanjiang Wu},
  \bibinfo{person}{Abhimanyu~Rajeshkumar Bambhaniya},
  \bibinfo{person}{Sarbartha Banerjee}, \bibinfo{person}{Tuhin Khare},
  \bibinfo{person}{Sudarshan Srinivasan}, \bibinfo{person}{Suvinay
  Subramanian}, \bibinfo{person}{Souvik Kundu}, \bibinfo{person}{Madhu Kumar},
  \bibinfo{person}{Midhilesh Elavazhagan}, \bibinfo{person}{William Won},
  {et~al\mbox{.}}} \bibinfo{year}{2026}\natexlab{}.
\newblock \showarticletitle{How Far Can Disaggregation Go? A Design-Space
  Exploration of Attention-FFN Disaggregation for Efficient MoE LLM Serving}.
\newblock \bibinfo{journal}{\emph{arXiv preprint arXiv:2605.28302}}
  (\bibinfo{year}{2026}).
\newblock


\bibitem[Xiao et~al\mbox{.}(2025)]%
        {xiao2025xdeepserve}
\bibfield{author}{\bibinfo{person}{Ao Xiao}, \bibinfo{person}{Bangzheng He},
  \bibinfo{person}{Baoquan Zhang}, \bibinfo{person}{Baoxing Huai},
  \bibinfo{person}{Bingji Wang}, \bibinfo{person}{Bo Wang}, \bibinfo{person}{Bo
  Xu}, \bibinfo{person}{Boyi Hou}, \bibinfo{person}{Chan Yang},
  \bibinfo{person}{Changhong Liu}, {et~al\mbox{.}}}
  \bibinfo{year}{2025}\natexlab{}.
\newblock \showarticletitle{xdeepserve: Model-as-a-service on huawei
  cloudmatrix384}.
\newblock \bibinfo{journal}{\emph{arXiv preprint arXiv:2508.02520}}
  (\bibinfo{year}{2025}).
\newblock


\bibitem[Yang et~al\mbox{.}(2025)]%
        {Qwen3Technical}
\bibfield{author}{\bibinfo{person}{An Yang}, \bibinfo{person}{Anfeng Li},
  \bibinfo{person}{Baosong Yang}, \bibinfo{person}{Beichen Zhang},
  \bibinfo{person}{Binyuan Hui}, \bibinfo{person}{Bo Zheng},
  \bibinfo{person}{Bowen Yu}, \bibinfo{person}{Chang Gao},
  \bibinfo{person}{Chengen Huang}, \bibinfo{person}{Chenxu Lv},
  {et~al\mbox{.}}} \bibinfo{year}{2025}\natexlab{}.
\newblock \showarticletitle{Qwen3 technical report}.
\newblock \bibinfo{journal}{\emph{arXiv preprint arXiv:2505.09388}}
  (\bibinfo{year}{2025}).
\newblock


\bibitem[Yu et~al\mbox{.}(2022)]%
        {yu2022orca}
\bibfield{author}{\bibinfo{person}{Gyeong-In Yu}, \bibinfo{person}{Joo~Seong
  Jeong}, \bibinfo{person}{Geon-Woo Kim}, \bibinfo{person}{Soojeong Kim}, {and}
  \bibinfo{person}{Byung-Gon Chun}.} \bibinfo{year}{2022}\natexlab{}.
\newblock \showarticletitle{Orca: A distributed serving system for
  $\{$Transformer-Based$\}$ generative models}. In
  \bibinfo{booktitle}{\emph{16th USENIX symposium on operating systems design
  and implementation (OSDI 22)}}. \bibinfo{pages}{521--538}.
\newblock


\bibitem[Zhang et~al\mbox{.}(2025b)]%
        {SGLangAntGroup2025}
\bibfield{author}{\bibinfo{person}{Tianyu Zhang}, \bibinfo{person}{Peng Zhang},
  \bibinfo{person}{Yusong Gao}, {and} \bibinfo{person}{Yun Zhang}.}
  \bibinfo{year}{2025}\natexlab{b}.
\newblock \bibinfo{title}{Together with {SGLang}: Best Practices for Serving
  {DeepSeek-R1} on {H20-96G}}.
\newblock
  \bibinfo{howpublished}{\url{https://lmsys.org/blog/2025-09-26-sglang-ant-group}}.
\newblock


\bibitem[Zhang et~al\mbox{.}(2025a)]%
        {zhang2025janus}
\bibfield{author}{\bibinfo{person}{Zhexiang Zhang}, \bibinfo{person}{Ye Wang},
  \bibinfo{person}{Yumiao Zhao}, \bibinfo{person}{Jiayu Xiao},
  \bibinfo{person}{Qianjing Yang}, \bibinfo{person}{Xiangyu Wang},
  \bibinfo{person}{Jingzhe Jiang}, \bibinfo{person}{Qizhen Weng},
  \bibinfo{person}{Ruichuan Chen}, \bibinfo{person}{Shaohuai Shi},
  {et~al\mbox{.}}} \bibinfo{year}{2025}\natexlab{a}.
\newblock \showarticletitle{Janus: Disaggregating Attention and Experts for
  Scalable MoE Inference}.
\newblock \bibinfo{journal}{\emph{arXiv preprint arXiv:2512.13525}}
  (\bibinfo{year}{2025}).
\newblock


\bibitem[Zhong et~al\mbox{.}(2024)]%
        {DistServe2024}
\bibfield{author}{\bibinfo{person}{Yinmin Zhong}, \bibinfo{person}{Shengyu
  Liu}, \bibinfo{person}{Junda Chen}, \bibinfo{person}{Jianbo Hu},
  \bibinfo{person}{Yibo Zhu}, \bibinfo{person}{Xuanzhe Liu},
  \bibinfo{person}{Xin Jin}, {and} \bibinfo{person}{Hao Zhang}.}
  \bibinfo{year}{2024}\natexlab{}.
\newblock \showarticletitle{$\{$DistServe$\}$: Disaggregating prefill and
  decoding for goodput-optimized large language model serving}. In
  \bibinfo{booktitle}{\emph{18th USENIX Symposium on Operating Systems Design
  and Implementation (OSDI 24)}}. \bibinfo{pages}{193--210}.
\newblock


\bibitem[Zhu et~al\mbox{.}(2025)]%
        {MegaScaleInfer2025}
\bibfield{author}{\bibinfo{person}{Ruidong Zhu}, \bibinfo{person}{Ziheng
  Jiang}, \bibinfo{person}{Chao Jin}, \bibinfo{person}{Peng Wu},
  \bibinfo{person}{Cesar~A Stuardo}, \bibinfo{person}{Dongyang Wang},
  \bibinfo{person}{Xinlei Zhang}, \bibinfo{person}{Huaping Zhou},
  \bibinfo{person}{Haoran Wei}, \bibinfo{person}{Yang Cheng}, {et~al\mbox{.}}}
  \bibinfo{year}{2025}\natexlab{}.
\newblock \showarticletitle{Megascale-infer: Serving mixture-of-experts at
  scale with disaggregated expert parallelism}.
\newblock \bibinfo{journal}{\emph{arXiv preprint arXiv:2504.02263}}
  (\bibinfo{year}{2025}).
\newblock


\end{thebibliography}

\end{document}